\documentstyle[11pt,newpasp,twoside,epsf]{article}
\def\edcomment#1{\iffalse\marginpar{\raggedright\sl#1\/}\else\relax\fi}
\reversemarginpar

\begin{document}
\title{The Toronto Red-Sequence Cluster Survey: First Results}
 \author{Michael D. Gladders and H.K.C. Yee}
\affil{Department of Astronomy, University of Toronto, 60 St. George Street, Toronto, Ontario, M5S 3H8, Canada}

\begin{abstract}
    The Toronto Red-Sequence Cluster Survey (TRCS) is a new galaxy
    cluster survey designed to provide a large sample of optically
    selected $0.1 < z < 1.4$ clusters. The planned survey data is 100
    square degrees of two color ($R$ and $z'$) imaging, with a 5$\sigma$
    depth $\sim$2 mag past $M^*$ at $z=1$. The primary scientific
    drivers of the survey are a derivation of $\Omega_{m}$ and $\sigma_8$ 
    (from $N(M,z)$ for clusters) and a study of cluster galaxy
    evolution with a complete sample. This paper gives a brief outline
    of the TRCS survey parameters and sketches the methods by which we
    intend to pursue the main scientific goals, including an explicit
    calculation of the expected survey completeness limits.  Some
    preliminary results from the first set of data ($\sim$ 6
    deg$^2$) are also given. These preliminary results provide new
    examples of rich $z\sim1$ clusters, strong cluster lensing, and a
    possible filament at $z\sim1$.
\end{abstract}

\section{The TRCS}
\paragraph{}
The Toronto Red-Sequence Cluster Survey (TRCS) is a major new
observational effort designed to identify and characterize a large
sample of galaxy clusters to redshifts as high as $z\sim1.4$. When
completed, the TRCS will be the largest imaging survey ever completed
on 4m telescopes, and will provide a large and homogeneous sample of
galaxy clusters for detailed follow-up study. The basic survey is
envisioned as 100 deg$^2$ of 2 filter ($R$ and $z'$) imaging, to a
depth which is $\sim$2 mag past $M^*$ at $z=1$ in both filters. The
design of the survey is based on a new method for identifying galaxy
clusters (Gladders \& Yee  2000a) developed specifically for the TRCS.
In brief, this method searches for clustering in the 5-D space of: x-y
positions, $R-z'$ color, $z'$ mag, and morphology in the form of a
concentration index. The x-y positions provide the surface density
enhancement. A color slice in the color-mag plane provides separation
in $z$ space via the {\it red sequence} of early-type galaxies in
clusters (Figure 1) and increases the S/N of density enhancements.
Morphology allows us to key onto early-type galaxies, the primary
population in cluster centers.

\begin{figure}
\plotone{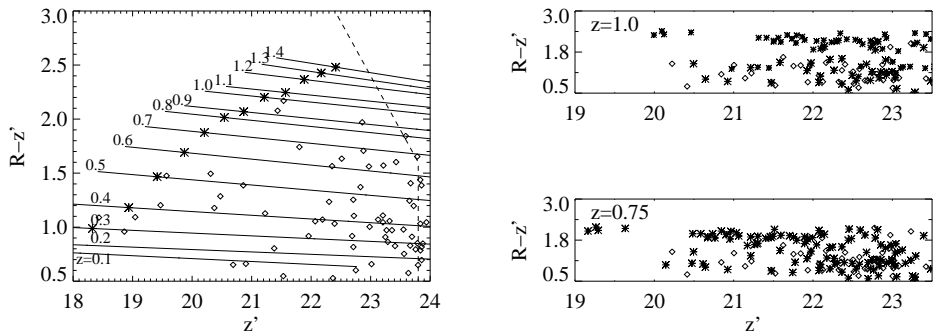}
\caption{
Modeled cluster CMDs to z=1.4 (left panel, solid lines), from Kodama
\& Arimoto (1997).  Diamonds indicate simulated field galaxies for a 1
arcmin$^2$ FOV. The *s show M$^{*}$ for each redshift. The TRCS
photometric completeness limits are shown (dashed line). We also show
real CMDs (right) for CL1322+3114 at $z=0.75$ and k-corrected to
$z=1$. These data are from HST images degraded to TRCS seeing and
depth.  Cluster image objects (*) and field image objects ($\diamond$)
are shown. Note the visibility of the cluster red sequence.}
\end{figure}
 
\subsection{Scientific Goals}
\paragraph{}
The TRCS is being driven by two major scientific goals. The first is
based on the theoretical prediction that the evolution of the
mass-spectrum of galaxy clusters with redshift, $N(M,z)$, should be a
strong function of two cosmological parameters, $\Omega_{m}$ and
$\sigma_8$ (Figure 2). The goal is to use the clusters identified in
the survey to measure $N(M,z)$ directly from the survey data. Redshift
can be estimated from the color of red sequence (e.g., L\'{o}pez-Cruz
\& Yee 2000), and the mass of each cluster can be estimated from its
richness, as measured by the parameter $B_{gc}$ (e.g., Yee \&
L\'{o}pez-Cruz 1999). The second major scientific goal is a study of
the cluster galaxy populations, which can be done using the TRCS for
the first time with a complete sample. The definition of a complete,
or volume limited, sample is derived from extensive simulations of the
survey selection functions.

\begin{figure}[htp]
\plotfiddle{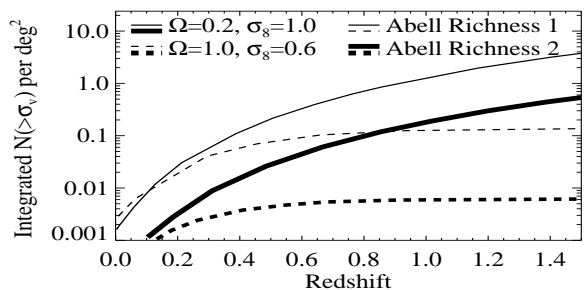}{3cm}{0.0}{50}{40}{-165}{-40}
\caption{The expected cumulative counts of clusters per~deg$^2$ for two
cosmologies, for Abell Richness Class (ARC) 1 and 2.  The 100\%
completeness redshift for ARC 1 is $\sim1.1$, and $\sim1.3$ for the
richer, rarer ARC 2 clusters.}
\end{figure}

\subsection{Survey Completeness and Selection Functions}
\paragraph{}
Any detailed understanding of the cosmological or galaxy evolution
results deriving from the TRCS requires a good understanding of the
survey selection functions. Specifically, we wish to know how well the
cluster-finding algorithm finds clusters of various sorts (as
described by various parameters). To this end, we have constructed a
number of cluster and field simulations (Gladders \& Yee 2000b) to
directly test the algorithm. A large suite of possible clusters have
been tested; the parameters describing the clusters are given in Table
1. The results of this process demonstrate that the TRCS should be
complete for all reasonable clusters of Abell Richness Class $\geq$ 1
clusters ($\sigma_{v}\geq750$ km s$^{-1}$) to at least $z=1.1$.
\begin{table}
\begin{tabular}{lll}
\hline
Parameter & Model Values & Notes \\\hline
LF $R$-band $M^*$ & -22.5, -22.25, -22.0 & $\alpha=-1.0$\\
Abell Richness counts & 35,44,56,72,93,120& Richness Classes 0-2 \\
NFW core scale radius & 0.1,0.2,0.3,0.4,0.5 & in $h^{-1}$ Mpc\\
ellipticity &0.0,0.2,0.4,0.6,0.8 & measured at 1 $h^{-1}$ Mpc \\ 
blue fraction & 0.1,0.5,0.65,0.8,0.9 &  \\
red sequence age & 9,10,11,11.5 & lower limit of SF in Gyr  \\
scatter in formation ages& 0.5,1.0,max & tophat width in Gyr  \\
cluster redshift&0--1.4&\\\hline
\end{tabular}
\caption{Cluster model parameters used to test the cluster finding 
algorithm as applied to the TRCS.}
\end{table}

\section{Some First Results}
The first run for the TRCS occurred at CFHT in May, 1999. A total of
21 pointings were acquired with the CFH12K camera, with each pointing
covering 0.272 deg$^2$. The bulk of the images have seeing better than
0\farcs7, with some as good 0\farcs5. At the time of writing, the
total TRCS dataset consists of about 35 deg$^2$ of data. However, the
results presented below are based on only the first 6 deg$^2$. Figure
3 shows a rich ($B_{gc}\sim2000$), compact cluster at photometric
redshift of $z\sim0.95$ (left panel). The cluster appears to be
embedded in a large
\begin{figure}[htp]
\plotfiddle{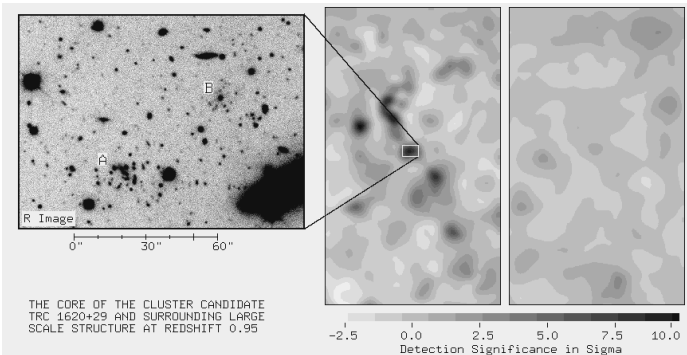}{3.65cm}{0.0}{48}{48}{-140}{-12}
\caption{The image on the left shows the core of the cluster (A) and a possibly associated
group (B).  The central and right panels show the surface density of
galaxies in a {\it much larger region} (roughly 20'x30') with and
without a color cut.}
\end{figure}
\noindent 
filamentary structure ($\sim$10 h$^{-1}$ Mpc long) traced out by
red galaxies (center panel). The excess is undetectable without a
color cut (right panel). The efficacy of color information in
isolating high-$z$ structures is obvious.

\paragraph{}
Figure 4 shows the cores of several other rich clusters, one of which
appears to have a gravitational arc. The success of the TRCS
in finding rich, high-z cluster candidates in the few
degrees searched so far implies that the total survey will contain
several hundred $z\geq0.8$ cluster candidates, a preliminary result
which is supportive of a low-density, high normalization cosmological
model. 

\begin{figure}
\plottwo{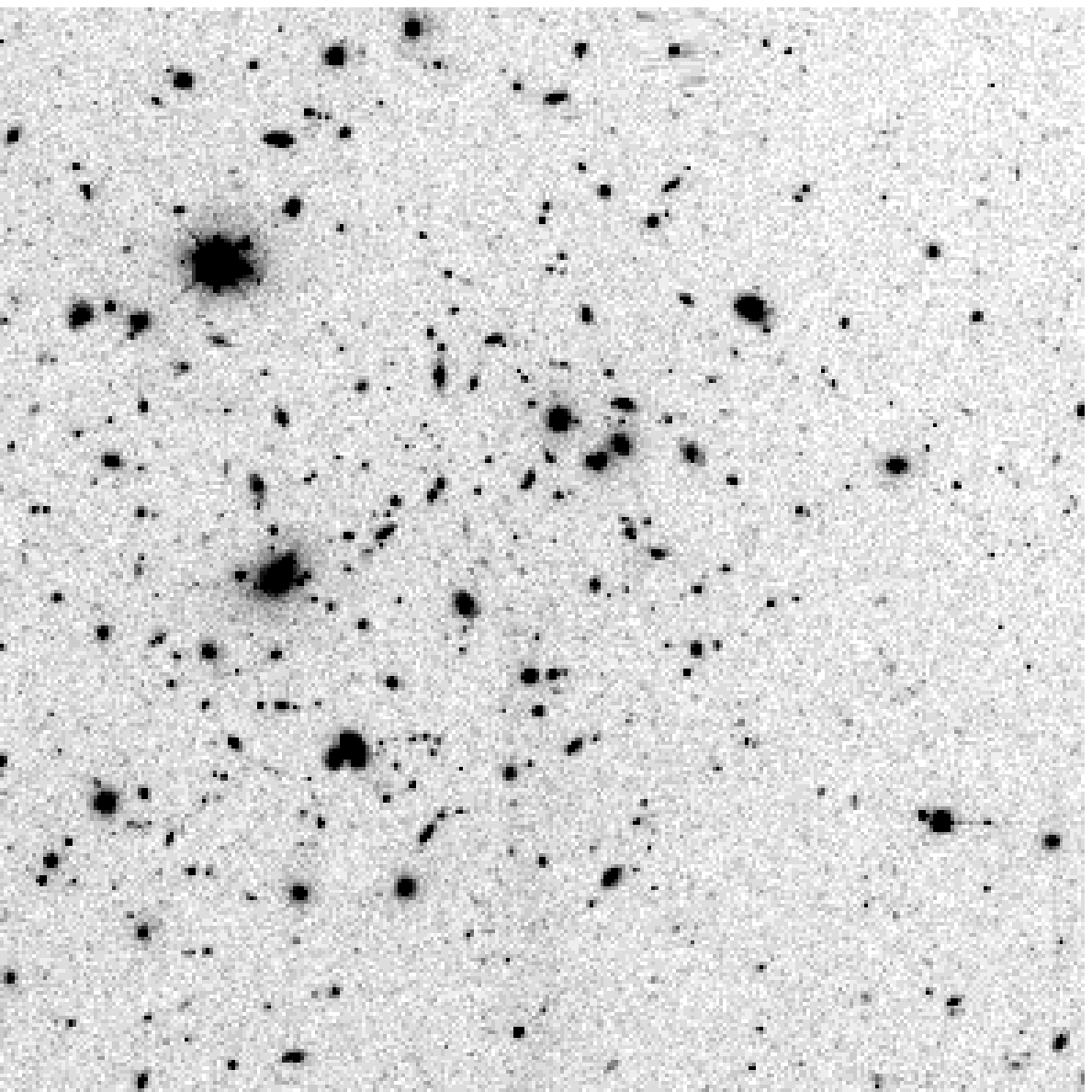}{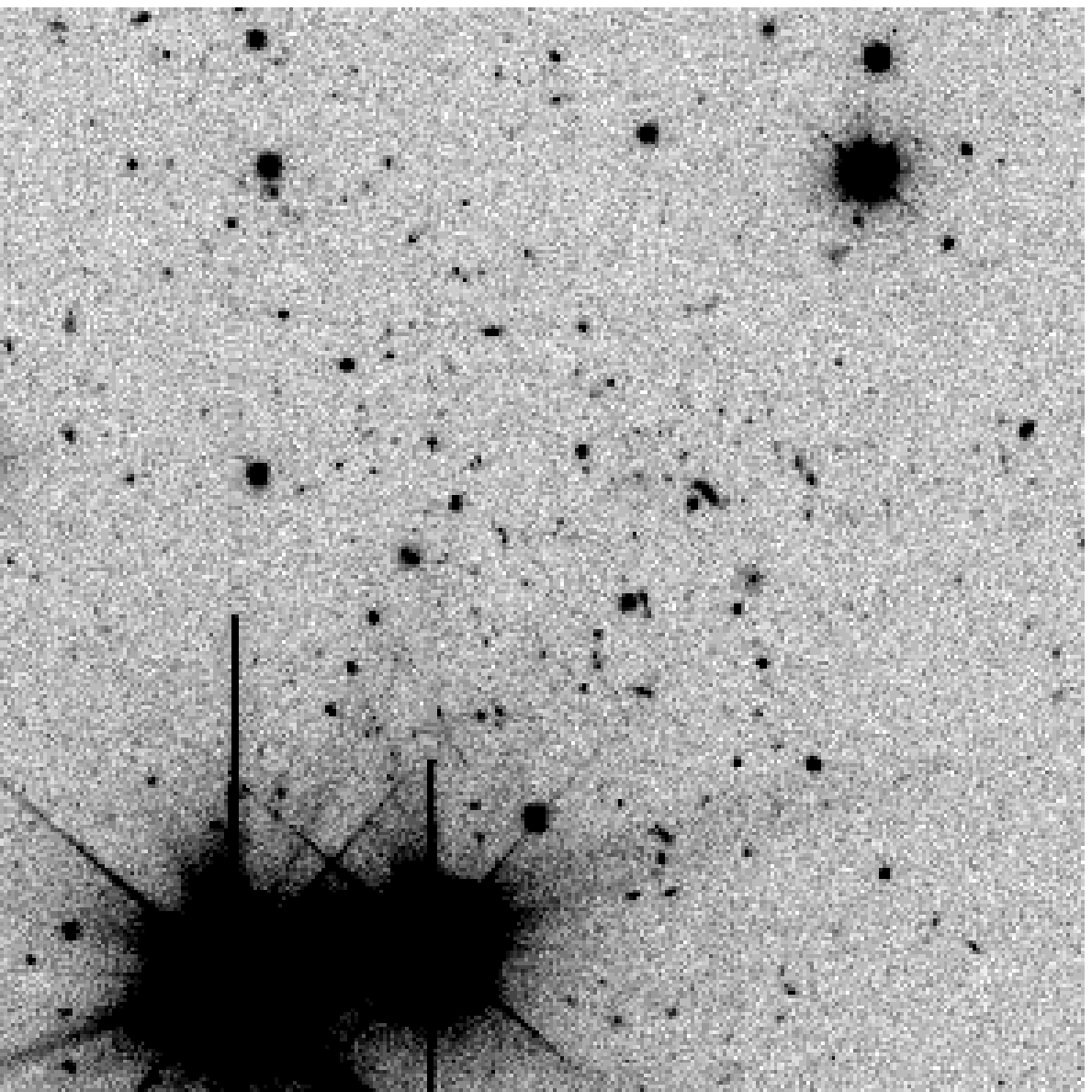}
\caption{
Two rich cluster candidates with estimated redshifts of 0.45 and 0.85
(left to right).  Though not readily visible here, the cluster at 0.85
appears to have a gravitational arc, consistent with its richness of
$B_{gc}\sim2500$. Large color images of these clusters can be found at
\texttt{http://www.astro.utoronto.ca/$\sim$gladders/TRCS/trcs1.html}}
\end{figure} 

\section{Secondary Projects}
The TRCS dataset is also well suited for a number of other detailed
studies. For example, preliminary work has already revealed a
significant population of extremely red ($R-z' \geq 3.5$) point
sources. Such objects are likely L and T dwarfs, with some
contamination by $z\geq5.5$ QSOs. Other studies are possible using the
survey data, e.g. studies of cluster lensing (strong and weak), cosmic
shear, halo structure, low surface brightness galaxies and early-type
galaxy correlations.

\references{
\reference Gladders, M.D., \& Yee, H.K.C. 2000a, in prep.
\reference Gladders, M.D., \& Yee, H.K.C. 2000b, in prep.
\reference Kodama, T., \& Arimoto, N. 1997, 320, 41
\reference  L\'{o}pez-Cruz, O., \& Yee, H.K.C. 2000, (to be submitted to ApJ)
\reference Yee, H.K.C., \& L\'{o}pez-Cruz, O. 1999, AJ, 117, 1985

}

\end{document}